\documentclass{nature}
\usepackage{amssymb,amsmath,epsfig}

\def\mnras{{\it Mon. Not. R. Astron. Soc.}}
\def\apj{Astrophys. J.}
\def\apjs{Astrophys. J. S.}

\title{Hydrogen 21-cm Intensity Mapping at redshift 0.8}

\author{Tzu-Ching Chang$^{1,2}$, Ue-Li Pen$^2$, Kevin Bandura$^3$, \& Jeffrey B. Peterson$^3$}

\begin{document}
\date{}
\maketitle
\addtolength{\topmargin}{0.5in}

\begin{affiliations}
\item IAA, Academia Sinica, P.O. Box 23-141, Taipei 10617, Taiwan  
\item CITA, University of Toronto,
  60 St.George St., Toronto, ON, M5S 3H8, Canada
\item  Department of Physics, Carnegie Mellon University,
  5000 Forbes Ave, Pittsburgh, PA 15213, USA
\end{affiliations} 

\begin{abstract}

Observations of 21-cm radio emission by neutral hydrogen at redshifts
z $\sim 0.5$ to $\sim 2.5$ are expected to provide a sensitive probe
of cosmic dark energy \cite{2008PhRvL.100i1303C}
\cite{2008PhRvL.100p1301L}. This is particularly true around the onset
of acceleration at z $\sim 1$, where traditional optical cosmology
becomes very difficult because of the infrared opacity of the
atmosphere. Hitherto, 21-cm emission has been detected only to z=0.24
\cite{2007MNRAS.376.1357L}. More distant galaxies generally are too
faint for individual detections but it is possible to measure the
aggregate emission from many unresolved galaxies in the 'cosmic
web'. Here we report a three dimensional 21-cm intensity field at
z=0.53 to 1.12. We then co-add HI emission from the volumes surrounding
about ten thousand galaxies (from the DEEP2 optical galaxy redshift
survey \cite{2001defi.conf..241D}). We detect the aggregate 21-cm glow
at a significance of $\sim 4$ sigma.

\end{abstract}

The DEEP2 optical redshift survey \cite{2001defi.conf..241D} provides
a rich sample for study of the Universe at redshifts near one.  The
team recorded optical spectra for 50,000 faint galaxies out to
redshift $z = 1.4$, utilizing the DEIMOS spectrograph on the Keck II
telescope.  The DEEP2 sample consists of four survey fields, each
$\sim 120' \times 30'$ in size. Roughly 10,000 of the DEEP2 galaxies
fall in the redshift range which overlaps the range of the radio data
described below.

Using the Green Bank Telescope (GBT), we recorded radio spectra across
two of the DEEP2 fields. We are searching for redshifted HI 21-cm
emission, and for such emission the radio spectra span the redshift
range $0.53 < z < 1.12$. This corresponds to a comoving distance
$1400$ to $2600$ h$^{-1}$ Mpc assuming a $\Lambda$CDM cosmology
\cite{WMAP52009ApJS..180..330K}.  The full-width-half-maximum GBT beam
is about 15', which corresponds to 9 h$^{-1}$ comoving Mpc at $z=0.8$
so the angular resolution falls far short of that needed to separate
one galaxy from the next.  In the redshift dimension the spatial
resolution is much higher than the transverse resolution. To keep the
data set manageable we average in frequency to a resolution 430kHz, or
2h$^{-1}$ comoving Mpc.  Unlike the optical survey we make no attempt
to detect individual galaxies with the radio data.  More observational
details can be found in the Supplementary Information Section 1.

We have two goals for the radio observations. First, we examine
whether it is possible to map the cosmic web in three dimensions,
without detecting individual galaxies, a technique called `intensity
mapping'. Second, we measure the 21 cm brightness and neutral gas
density at high redshift.  To accomplish the first goal we study the
21 cm data by itself. For the second goal we draw on the DEEP2 galaxy
positions as guides, using them to locate likely bright spots of the
21 cm glow. If sufficient gas is present, and if the intensity mapping
technique is viable, it will be possible to use this technique to
economically map over 50 times the volume that has so far been
surveyed.

To get to the 21 cm signal we must remove two much brighter sources of
flux from the data, Radio Frequency Interference (RFI) from
terrestrial transmitters and broadband (continuum) emission by
astronomical sources within and outside of the Milky Way.

\begin{figure}
\psfig{file=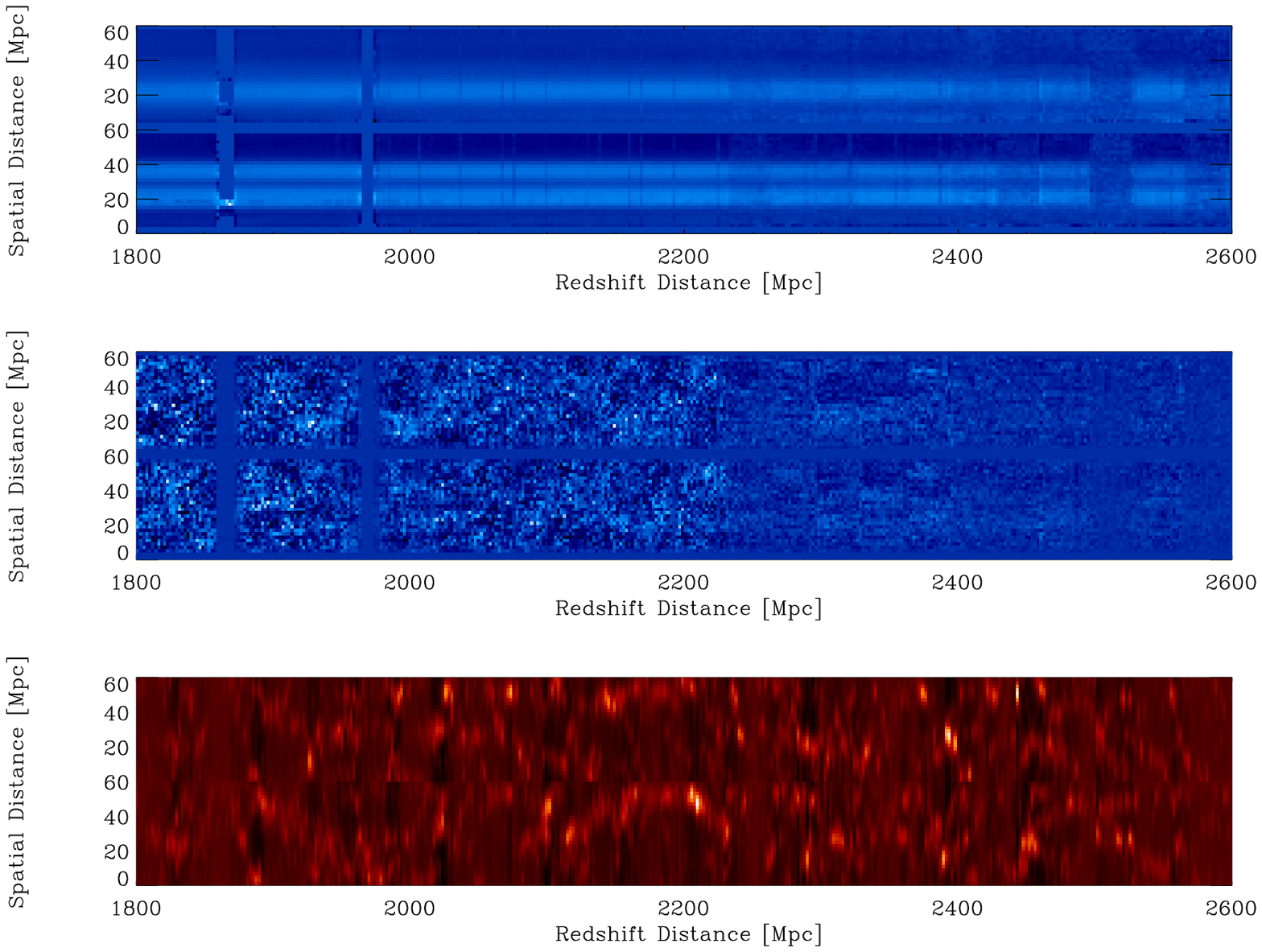, width=6in}
\caption{{\bf Spectra of DEEP2 Field-4.} The top two panels show radio
  flux arranged with redshift horizontal and right ascension vertical.
  Each panel contains data collected in two declination bins,
  separated by $15'$, roughly one GBT beam width. The higher
  declination strip occupies the top half of each panel. The pixels
  are 2 h$^{-1}$ comoving Mpc in size in each dimension.  The top
  panel shows measured flux after the polarization cut has removed the
  brightest terrestrial emission. The rms fluctuation of the map is
  128mK. Vertical structures in the top panel are due to residual RFI:
  the wide stripes are digital television signals and narrow vertical
  features are analog television carriers. Redshift windows free of
  RFI are rare on the right side of the plot, which corresponds to
  greater redshift and lower frequency. The horizontal bright stripes
  are due to continuum emissions by astronomical sources( NVSS
  J022806+003117 and NVSS J022938+002513), and the width of these
  stripes shows the GBT beam width.  The middle panel shows the
  inverse-variance-weighted radio brightness temperature, after
  subtraction of continuum sources.  The weighted rms fluctuation is
  3.8mK.  Even though the standard deviation of the flux values in
  this images has been reduced by more than a factor of 30 compared to
  the top panel, residual RFI and continuum emission dominates the
  overall variance. The bottom panel shows the optical galaxy density
  in the DEEP2 catalog, smoothed to match the resolution of the radio
  data. The rms fluctuation of the map is 1.8. The cross correlation
  function in Figure \ref {fig:corr} is calculated by multiplying the
  middle and bottom panels with a relative displacement (lag) in
  redshift, then calculating the variance of the product map.
    \label{fig:rfi} }
\end{figure}

We use polarization to identify and excise unwanted signals.
Television, mobile telephone transmitters, and other terrestrial
sources produce strongly polarized flux which remains polarized even
after scattering.  The GBT has two linearly polarized feed antennas at
the focus of the parabolic reflector.  RFI entering the feed antenna
sidelobes produces signals in the data stream that correlate between
the two polarization channels.  21-cm emission is not polarized, so we
calculate the cross-correlation coefficient between the two polarizations
and cut any flux with coefficient greater than 2\%.  This removes
$\sim 5\%$ of the data. The radio data, after this cut, is shown in
the top panel of Figure \ref{fig:rfi}. At this stage of the analysis
astronomical continuum sources present a fluctuation of brightness
temperature across the sky of rms amplitude $\sim 125$ mK, which is
about a thousand times larger than the HI signal.

We remove astronomical continuum flux using a new matrix-based method
\cite{2009MNRAS.399..181P}.  After a calibration procedure described
in the Supplementary Information we arrange the data into a matrix in
which the row index represents celestial coordinates, and the column
index represents the observation frequency (redshift). All three
panels in Figure \ref{fig:rfi} show such matrices.  Continuum sources,
including bright radio sources, blends of weak extragalactic sources,
and Galactic synchrotron sources, extend across all columns of the
matrix. They show up as horizontal stripes in the top panel of Figure
\ref{fig:rfi}. In contrast, 21-cm sources are tightly localized to a
few columns.  Within this matrix, a continuum source can be factored
into a product of a function of position (for an isolated point source
this is the beam pattern of GBT), and a low order function of
frequency (the smooth spectrum), $T(x,\nu) = f(x)g(\nu)$.  The
functions $f(x)$ and $g(\nu)$ are singular eigenmodes of the matrix,
so we perform a singular-value decomposition (SVD) to detect continuum
flux, which we subtract.  We are then left with the brightness
temperature field $\delta T_{b}$, shown in the middle panel of Figure
\ref{fig:rfi}.  The SVD technique is non-parametric. No particular
mathematical form for the spectra of the continuum sources is assumed.

We next examine the radio data without referring to the optical
survey.  After foreground subtraction, keeping only the component that
is consistent over several days, we find the intensity field has a
temperature fluctuation $464 \pm 277 \mu$K, on pixel scales of (2
h$^{-1} )^3$ comoving Mpc$^{3}$. Note that our resolution element, (9
h$^{-1} )^3$ comoving Mpc$^{3}$, is larger than the pixel. The noise
in this measurement exceeds that expected from the antenna temperature
and is likely due to residuals of emission by terrestrial transmitters
and errors in the astronomical continuum removal process. Because of
the weak statistical significance, this fluctuation amplitude should
be treated an upper limit to the 21-cm brightness auto-correlation,
rather than a detection of cosmic structure.

\begin{figure}
\centerline{\epsfig{file=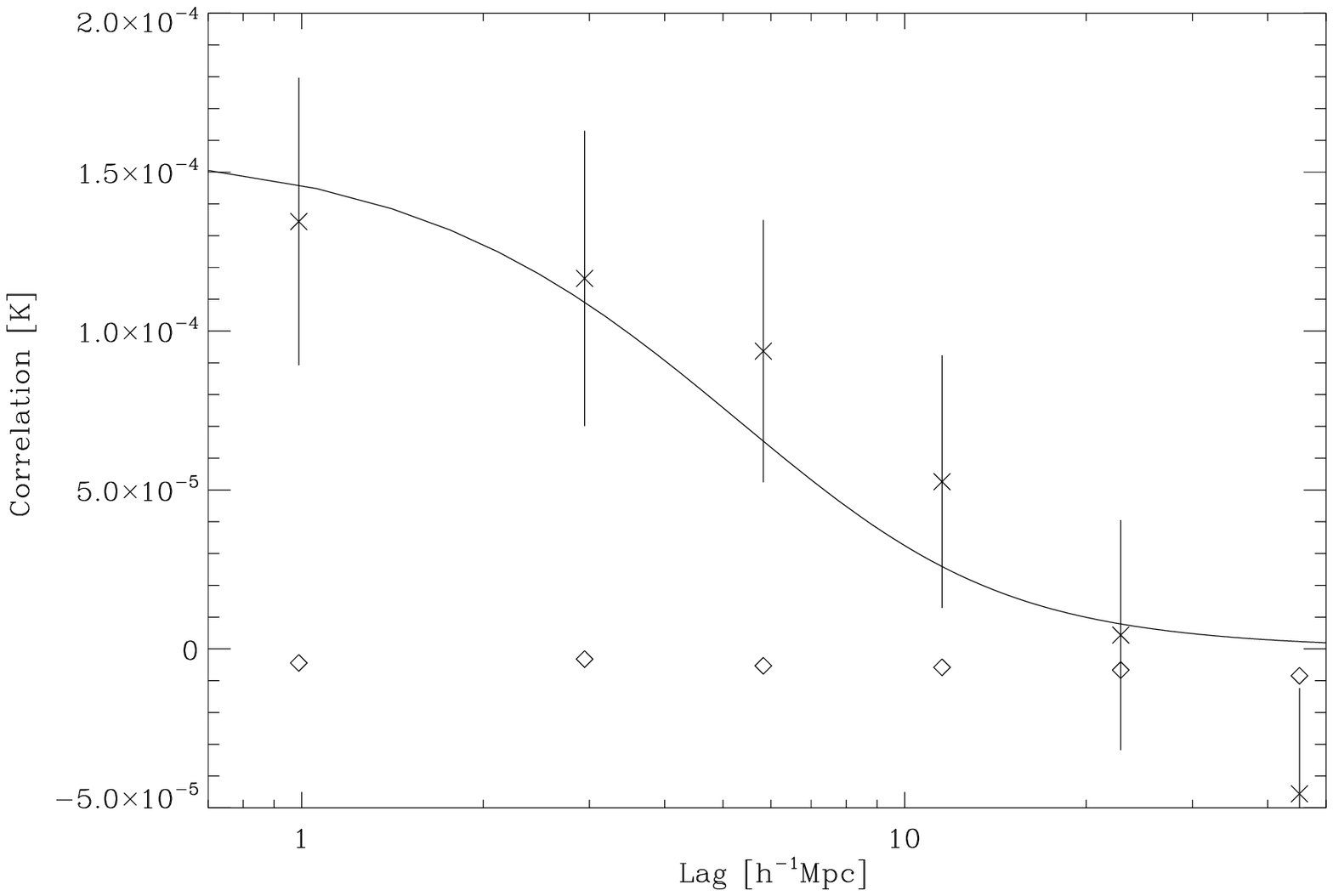, width=\columnwidth}}
  \caption{{\bf The cross correlation between the DEEP2 density field
      and GBT HI brightness temperature.} The crosses are the measured
    cross-correlation temperature, while error bars are the 1$\sigma$
    bootstrap errors generated using randomized optical data, and the
    diamonds are the mean null-test values over 1,000 randomizations
    as described in SI; the same bootstrap procedure performed on
    randomized radio data returns very similar null-test values and
    error bars. The solid line is a DEEP2 galaxy correlation model
    which assumes a power law correlation and includes the GBT
    telescope beam pattern as well as velocity distortions, and uses
    the best-fit value of the cross-correlation amplitude.
  \label{fig:corr}}
\end{figure}

To further reduce uncertainty we proceed with the cross-correlation
(stacking) technique. Cross-correlation reduces the error because
terrestrial RFI and residuals of the continuum sources are randomly
located compared to the locations of optically bright galaxies.  To
carry out the cross correlation we arrange the DEEP2 data in the same
matrix as the radio data, as shown in Figure \ref{fig:rfi} .  We
calculate the weighted cross correlation $\zeta $ between the data in
the bottom two panels of the figure, producing the correlation
function in Figure \ref{fig:corr}.  We detect significant
cross-correlation power out to lag $\sim 10$ h$^{-1}$ comoving Mpc.

To check this cross-correlation result we carry out a statistical null
test. We randomize the optical redshifts many times, each time repeating
the correlation calculation. We find no significant correlation in the
randomized sets and we use the bootstrap variance to estimate the
uncertainties in our measurements. The null test confirms that the
residual RFI and astronomical continuum sources are unlikely to cause
false detection of 21-cm emission.

The measured cross correlation can be compared to a model
prediction.  Locations of optically cataloged galaxies are known to be
correlated amongst themselves, and 21-cm emission is also thought to
originate in galaxies. We therefore model the cross-correlation by adopting the
DEEP2 optical galaxy auto-correlation power law
\cite{2004ApJ...609..525C} , $\xi(R)=(R/R_0)^{-1.66}$, where $R_0 =
3.53$ h$^{-1}$ Mpc at z=0.8, which we convolve with the telescope primary beam
in the transverse direction.  In the radial direction we must account
for peculiar velocities.  The pairwise velocity distribution is
modeled as a Gaussian with standard deviation $\sigma_{12}=395$
km/sec, using the relation $\sigma_{12} \sim
H(z)~R_0$~\cite{2004ApJ...609..525C} \cite{2003MNRAS.344..847M}. The
expected correlation, calculated using this model, is plotted in
Figure \ref{fig:corr} using the best fit value of the correlation
amplitude.

We use the correlation amplitude to constrain the
neutral hydrogen density at redshift 0.8.
The cross-correlation $\zeta $ between the optical galaxy density
field and the neutral hydrogen temperature field is related to the
density structure by (e.g., \cite{2008PhRvL.100i1303C})
\begin{equation}
\zeta = \left<\Delta T_b \delta_{opt}\right> = 284 ~b ~r ~\delta_{opt}^{2}  \left({\Omega_{HI} \over 10^{-3}}\right)\left({h \over 0.73}\right) 
\left({\Omega_m + (1+z)^{-3} \Omega_\Lambda\over 0.37}\right)^{-0.5}  
\left({1+z \over 1.8}\right)^{0.5} {\rm \mu K}.
\label{eq:cross}
\end{equation}
where $\Delta T_b$ is the neutral hydrogen 21-cm brightness
temperature fluctuations, $T_b = 284 ~\mu K$ is the 21-cm mean sky
brightness temperature at $z=0.8$, $\Omega_{HI} \equiv \rho_{HI}/\rho_{c,0}$ is the HI density over the present-day critical
density, $\Omega_m$ the matter density, and $h$ is the current
expansion rate in units of $100$ km/s Mpc$^{-1}$. $\delta_{opt}$ is
the optical density field, which is related to the neutral hydrogen
density field, $\delta_{HI} = {br} \delta_{opt}$, where $b =
\left<\delta_{HI}^2\right>^{1/2} / \left<\delta_{opt}^2\right>^{1/2}$
is the bias factor, and $r = \left<\delta_{HI}
\delta_{opt}\right>/(\left<\delta_{HI}^2\right>\left<\delta_{opt}^2\right>)^{1/2}$
is the stochasticity.  Note that $|r| \leq 1$, and our data show $r$
to be positive. The effective $\delta_{opt}^2$ values for DEEP2
Field-3 and Field-4 calculated with simulations described in
Supplementary Information are 2.3 and 3.3, respectively.  A cosmic
hydrogen fraction of $\rho_{H} / \rho_b = 0.75$ is assumed.

The correlation function at zero lag has a 21-cm brightness
temperature $157 \pm 42 \mu$K at a mean effective redshift of $z=0.8$,
from which we infer a value of $\Omega_{HI} = (5.5 \pm 1.5) \times
10^{-4} \times (1/rb)$.  Combining all data in Figure \ref{fig:corr},
the statistical significance of the detection is at the four sigma level.

The cross correlation technique measures only the 21-cm component that
clusters near optically bright galaxies.  There may be additional
neutral gas at high redshift that is more broadly distributed, or the
gas may be clumped, but at locations not near the DEEP2 galaxies.  If
so, our fitting of the optical galaxy correlation function to the data
underweights this component. Our detection should therefore be treated
as a lower bound to the total neutral gas density.

We estimate the contribution of the DEEP2 galaxies to the total
zero-lag 21-cm flux variance is $\sim 20\%$ of the total.  The radio
data has many galaxies in each independent resolution element so we
can not separate flux in the zero-lag bin due to the 21-cm emission of
the DEEP2 galaxies themselves from that due to aggregate emission of
other galaxies concentrated nearby.  The effective resolution element
in this survey is $(9 h^{-1}$ Mpc$)^3$, determined by the telescope
angular resolution and pairwise velocity dispersion.  On average there
are five DEEP2 galaxies in each resolution element. We assume these
have 21 cm luminosities similar to low redshift galaxies in estimating
their contribution.

At $z \sim 1$, the neutral gas density $\Omega_{HI}$ is particularly
difficult to measure. At low redshifts (using HST), $\Omega_{HI}$ has
been measured via the Lyman-alpha-line absorption of ultraviolet light
from distant quasars. Extrapolating to redshift 1.2, $\Omega_{HI}= 7.2
\pm 2.2 \times 10^{-4}$ is found\cite{2006ApJ...636..610R}. Similar
observations can also be made with ground based telescopes at
redshifts above two, since these wavelengths penetrate the
atmosphere. However at redshift $z=2.2$, a substantially lower neutral
hydrogen content $\Omega_{HI}= 3.9 \pm 0.7 \times 10^{-4} $is
found\cite{2009ApJ...696.1543P}. Despite the high redshift, the ground
based measurement is consistent with the measured present day value
\cite{2005MNRAS.359L..30Z}. This implies little evolution of the
average gas density, at conflict with the clear evolution of star
formation rate.  Our data constrain the combination $\Omega_{HI} rb $,
but there is so far no observational constraint on the 21-cm
stochasticity $r$ or the bias $b$.  Theoretical estimates of $rb$ lie
in the range 0.5 to 2, a range too wide for the 21 cm data to weigh
in. With further 21 cm observations $r$ can be measured by detecting
both the auto and cross correlations, and $b$ can be determined by
measuring velocity space distortions\cite{2010arXiv1001.4811M}. This
in turn would allow measurement of the neutral gas density at $z \sim
1$ via 21 cm intensity mapping.

\begin{addendum}
\item We are grateful to the GBT support staff, in particular Toney
  Minter and Paul Ruffle, for their generous help with the
  observation.  We thank Kevin Blagrave, Olivier Dor\'e, Patrick
  McDonald, Jonathan Sievers, Kris Sigurdson, Renbin Yen for many
  useful discussions.  We acknowledge financial support by NSERC, NSF,
  and NRAO.  The National Radio Astronomy Observatory is a facility of
  the National Science Foundation operated under cooperative agreement
  by Associated Universities, Inc.
\item[Author Contribution] T.-C.C. and U.-L.P. analyzed and
  interpreted the data. K.B. conducted the remote
  observations. J.B.P. was in charge of the paper writing. All authors
  were present at the telescope for the on-site observations and
  contributed to the writing of the manuscript and Supplementary
  Information.
\item[Competing Interests] The authors declare that they have no
competing financial interests.
\item[Correspondence] Correspondence and requests for materials
should be addressed to T.-C. Chang.~(email: tchang@cita.utoronto.ca).
\end{addendum}

\newpage

\newpage
\newpage

\end{document}


\date{}
\maketitle
\addtolength{\topmargin}{0.3in}

\section{Observation details}

We use the Green Bank Telescope(GBT) in drift scan mode to record spectra
on Fields 3 and 4 of the DEEP2 survey area.  We park the telescope and
let the sky drift through the beam of GBT.  Each scan takes about six
minutes. The telescope is then moved to the leading edge of the field at
a new declination and parked, again letting the sky drift across
the beam.  Six such scans are used to cover the field at a series of
declinations separated by 0.05$^{\circ}$. Since the GBT beam size is
approximately 0.25$^{\circ}$ at these frequencies, the scan pattern
creates a raster of overlapping beams.  Averaged spectra for each
polarization are saved every 0.5 seconds. We recorded a total of 10.5
hours of integration time on Field-3 and 4.8 hours on Field-4. A noise
source at the feed point is switched on and off every 0.5s, increasing
the system temperature by 10\% while on.  We find gain fluctuations cause
apparent flux changes over the drift scan period that exceed real sky
signal variations. Rapid calibration using the noise source is required
to measure sky flux variations, which are used in the sky maps shown in
this paper.  The noise source is only used to calibrate the average gain,
not the spectral structure.  Since our signal is spectral in nature, the
low level of the noise source does not introduce substantial additional
noise in the map.

\section{Radio Frequency Interference (RFI) Removal} 

Although there is a radio quiet zone in the immediate vicinity of GBT,
in the surrounding area these frequencies are used for television
transmission as well as mobile telephone service.  These terrestrial
sources are single-mode and therefore 100\% polarized at the source.
The GBT has two linearly polarized feed antennas at the prime focus of
the parabolic reflector.  The feed system is about 150 meters above
the ground, so high that flux from transmitters outside the radio
quiet zone enters the sidelobes of the two feed antennas.  This
produces signals in the data stream that strongly correlate between
the two polarization channels. Some astronomical sources are partially
polarized, particularly pulsars, however 21-cm emission is not
polarized. We therefore use polarization as a discriminating factor
allowing identification and removal of terrestrial flux, to which we
give the name Radio Frequency Interference (RFI).

For each time and
frequency we calculate the inter-feed cross-correlation coefficient
$q$, which is the ratio of the cross-polarized flux to the unpolarized
flux. We flag as RFI, and discard, data that has $q > 0.02$. This
procedure removes $\sim 5\%$ of the data. The $q$ cut removes RFI that
arrives at the feed at high signal to noise ratio, however lower level
RFI remains.  While this residual RFI raises the noise level in the
DEEP2-21-cm cross-correlation there is no reason to think that DEEP2
galaxy redshifts are correlated with the frequencies of terrestrial
television transmitters and the low-level RFI should therefore not
provide a spurious apparent cross-correlation signal. The success of
the null test described below shows the absence of such spurious
correlation.

In addition to the $q$ cut we subtract the mean flux in each spectral
channel. The assumption here is that any flux localized in frequency,
but present for all telescope pointings, is likely terrestrial.

\section{Calibration}

The flux calibration is carried out by observing the bright radio
source 3C48.  Alternating the telescope pointing on source and then
off source allows measurement of the response to 3C48 compared to the
noise source, which is continuously switching throughout. The presence
of 3C48 in the beam changes the total system temperature by a factor
of two, which allows an accurate calibration while avoiding error due
to non-linearities in the correlator sampling system.

\section{Foreground (Continuum) Removal} 
\label{sec:svd}

At these frequencies the primary broadband emission mechanism in
astronomical sources is widely accepted to be synchrotron radiation
by relativistic electrons. This mechanism generically produces a very
smooth spectrum with at most a gentle slope across our frequency
range. Sources in the Milky Way, as well as extragalactic sources
contribute to the continuum flux and our procedure treats all such
sources as equivalent. We call the composite broadband flux
`foreground' emissions even though studies of extragalactic radio
sources typically show that many have redshift greater than one.

In contrast to the foregrounds, the 21-cm brightness should be highly
structured in its spectrum. However, the variation from pixel to pixel
due to the foregrounds is three orders of magnitude larger than the
21-cm signal, so it is challenging to calibrate the spectral response
with sufficient accuracy, especially for an instrument that has time
variable gain. Fortunately, we find that instrumental gain
fluctuations are primarily achromatic, and any chromatic variation we
do see are fit by low order polynomials in frequency.  We are
searching for 21cm signal fluctuations that span few-percent frequency
ranges. The spectral stability provided by the noise source is
sufficient across such a span.  Often one subtracts a low order
polynomial in frequency to remove the tendency of drift of the
spectral calibration to create foreground residuals.  However, since
the RFI cutting generates a complicated window structure in the
spectrum, we instead developed a novel non-parametric technique based
on singular-value-decomposition (SVD).

We arrange the data into a matrix in which the row index represents
celestial coordinates of the telescope pointings and the column index
represents the observation frequency (redshift). All three panels in
Figure 1 of the main text show such matrices.  Continuum sources
extend across all columns of the matrix.  In contrast, 21-cm sources
are tightly localized to a few columns.  Continuum sources can be
factored into a product of a function of position and a low order
function of frequency. That is, we assume the brightness temperature
$T$ is a separable function of space $x$ and frequency $\nu$,
$T(x,\nu) = f(x)g(\nu)$. Then the product functions $f(x)$ and
$g(\nu)$ are singular eigenmodes of the matrix.  We perform a
singular-value decomposition (SVD) on the matrix to find these modes.
The dominant right eigenvector $g_1(\nu)$ provides a correction to the
gain calibration in this model.  For each raster, we tag the ten
eigenmodes with the largest singular eigenvalues as contribution from
the foregrounds, accounting for a spectrally smooth component and
variations to that. We remove these eigenmodes. To this point the
procedure is carried out on individual drift scan rasters, matrices
arranged as a function of drift scan time (spatial sampling) and redshifts. Next
we accumulate all data for each field into a six-declination spatial
and redshift map.  Large-spatial-scale fluctuations remain in this
data, so we remove three more SVD modes.  We are then left with the
brightness temperature field $\delta T_{b}$, shown in the middle panel
of Figure 1.

Visual inspection of the foreground subtracted maps indicates that
the residual arises from time and frequency variable RFI, which
does not take the factorization form that we assumed for the foregrounds.
Work is in progress to test active noise cancellation, which may allow
for cleaner foreground subtraction.

In principle, one might also worry about frequency dependence of telescope
beams.  The observations were scanned at very fine spatial sampling,
which would allow us to degrade the resolution at all frequencies to the
lowest common denominator.  Unfortunately, the foreground residual
appear RFI dominated, so the beam compensation was not implemented.
 
\section{Cross Correlation}
\label{sec:cross}

We calculate the cross-correlation between the HI and optical
three-dimensional density fields, as a function of separation in the
redshift direction.  The DEEP2 density, $\rho_{opt}$, is
estimated directly from the DEEP2 catalog; galaxies are binned into a
cube with a 3D cell size of $3' \times 3' \times 2$ h$^{-1}$ Mpc,
according to their 3D optical positions. For every 370 h$^{-1}$ Mpc in
the redshift direction, the mean 3D optical galaxy density $\bar{\rho}$ is
computed.  The optical density contrast is then given by $\delta_{opt}
= \rho_{opt} / \bar{\rho} -1$.  We further subtract off the mean value
of $\delta_{opt}$ in each redshift interval, to match the analysis of
the radio data cube. This optical density field $\delta_{opt}$ is further
convolved with the GBT primary beam response, a Gaussian function with
FWHM of $15'$. This is shown in the bottom panel of Figure 1. The
cross-correlation signal is then the weighted sum of the products of
the two fields, $\zeta(d) = \sum_{x,z} \left[ w_{opt}(x) ~w_{HI}(x)
  ~w_{HI}(z+d) ~\delta_{opt}(x,z) ~\delta_{HI}(x,z+d) \right] / \sum_{x,z} 
\left[w_{opt}(x) w_{HI}(x) w_{HI}(z+d) \right] $, summed over all
spatial $x$ and redshift $z$ pixels, where $w_{HI}(x)$ is given by the
number of observations in a spatial pixel, summed over all redshifts.
$w_{HI}(z)$ is given by the inverse-variance of $\delta_{HI}(x,z)$ at a
given redshift bin, and $w_{opt}(x)$ is obtained by collapsing the
redshift direction such that the 2D weights represents the optical
survey masks, and $d$ is the separation in redshift space.
The weighting functions effectively put our cross-correlation
measurement at mean redshift $z=0.8$.

At our spatial resolution, there are 10 NVSS radio sources per beam.
Most of these radio sources are AGN's; in comparison, there are 500
DEEP2 galaxies per beam.  It is really the redshift resolution that
results in any meaningful signal, and we can see that the correlation
indeed drops to zero at large lags, meaning we are not measuring the
correlation of AGN continuum emission with galaxies.  Radio continuum
emission is not correlated spectrally.

\section{Auto Correlation}
\label{sec:auto}

We attempted to measure the auto-correlation of the HI temperature
field after the RFI and foreground removal. Our foreground-subtracted
maps contain residual RFI, which has a time variable component. In
addition, the (time variable) RFI cuts affect the quality of our
measurement of the gain versus frequency, and therefore limit the
precision of foreground removal. The 21-cm signal should however be
constant from day to day. We therefore calculate the auto-correlation
of a given field by cross-correlating radio data obtained on different
observing dates. The correlation values calculated this way are lower
than the rms of the intensity map shown in Figure 1.  The mean HI
auto-correlation of Field-3 and Field-4 is measured to be $464 \pm 277
\mu$K, which should be interpreted as an upper limit.

The cross-correlation technique can only find 21-cm concentrations
that happen to cluster around optically bright galaxies. The auto
correlation, if detected, would additionally include 21-cm structures
that are {\it not} near bright galaxies. In other words the
stochasticity $r$ (in Eqn. (1) in the main text) is less than one and we
expect the auto-correlation to be larger than the
cross-correlation. The auto-correlation uncertainty is about a factor two
larger than the cross-correlation signal, so despite the RFI and gain
fluctuations at GBT, we come very close to an auto-correlation
detection. This is important because it implies that using a 100 m
telescope at a quieter site, it should be possible to detect high
redshift 21-cm structure directly, without use of an optical redshift
survey catalog.

\section{Error Estimation and Null Tests} 

To study the error in our determination of the cross-correlation we
test the (null) hypothesis that there is no real correlation in the
data.  This requires generating a large ensemble of random data sets with the
same statistical distribution as the data from the sky.  Since this
distribution is not {\it a priori} known, we need to estimate it from
our data.  Bootstrap sampling provides such a procedure.

The bootstrap error is calculated as follows: for every 3D position we
assign an HI brightness temperature in the redshift direction drawn
randomly from the radio cube, generating an example of a randomized HI
temperature field.  Repeating this process many times, the resulting
cross-correlation values serve as null tests and the distribution an
error estimator.  The optical galaxy angular position information is
retained in this exercise, as are the optical survey masks.

We repeat this procedure, this time keeping the HI temperature field
fixed and randomizing the optical density field in the redshift
direction, retaining the optical survey masks. We find for both
versions of the randomization the null tests succeed, and the
variances in the cross-correlation values are very similar with 1,000
random samplings; these are used to generate our error estimate on the
cross-correlations, and the null tests values are plotted in Figure 2
in the main text.

\section{Adjustment for 21-cm signal loss}
\label{sec:sim}

To determine how much the foreground removal process attenuates the
21-cm signal, we generate a simulated 21-cm sky. We start with the
DEEP2 density field, smooth it with the GBT beam pattern as described
in Section~\ref{sec:cross}, and multiply it by 21-cm brightness
temperature 100$\mu$K. 

First, we cross-correlate this simulated 21-cm field with the optical
density field using all the weighting functions derived from the data,
as described in Section~\ref{sec:cross}. Because both parts of the
correlation product contain the field $\delta_{opt}$ the calculation
yields the effective (weighted) variance $\delta_{opt}^2$ of each of
the DEEP2 fields.  We use this to solve for the normalized correlation
function in Figure 2 as well as the mean hydrogen density.  We use
this procedure because it accounts for the complex weighting functions
produced by RFI cuts, inverse-variance weighting, and the optical
survey masks.  The effective $\delta_{opt}^2$ calculated using this
procedure for Field-3 and Field-4 are 2.3 and 3.3, respectively.

To measure the loss of 21-cm signal due to the SVD process we add the
simulated 21-cm field to the original drift-scan-raster radio data and
pass this through the entire analysis process. After foreground
removal using the SVD procedure described in Section~\ref{sec:svd}, we
calculate the cross-correlation of the simulation-added data,
$\zeta(d+s)$, where $d$ denotes actual radio data and $s$ the
simulation, and compare to the cross-correlation of data alone,
$\zeta(d)$. We compute the retention factor $h=
\left[\zeta(d+s)-\zeta(d)\right]/\zeta_\circ(s)$ as a function of
subtracted SVD modes. Here $\zeta_\circ(s)$ is the cross-correlation
of the simulated 21-cm field without going through the SVD process ,
while both $\zeta(d+s)$ and $\zeta(d)$ experience the SVD process.

We find the signal loss increases with the number of
subtracted SVD modes, as expected, although after the first few SVD
modes both the correlation values and the signal loss level off with
number of SVD modes, suggesting a converging value for the
cross-correlation. For the 13 SVD modes we settled on, the loss is
$\sim 30\%$ and we scale the correlation signal and error accordingly.

\section{Volume and Mass Scales}

We recast our results in units of hydrogen mass, which may be more
familiar to some than brightness temperature. The GBT angular
resolution, along with the galaxy pairwise velocity dispersion, define
an effective resolution volume element $dV=(9 h^{-1} {\rm Mpc})^3$.
The whole survey, including both fields, has 2000 such volume
elements. These cells contain on average five DEEP2 galaxies since
there are about 5000 galaxies per DEEP2 field. The total volume is
about $126\times 18\times 740 (h^{-1} {\rm Mpc}^3$).

The average optical luminosity per volume element is $L=V j_B =
3.5\times 10^{11} L_{\odot}$ \cite{2006ApJ...647..853W}.  The galaxies
in the DEEP2 catalog account for more than half the light in this
wavelength range.  Integrating the best fit luminosity function for
the optical galaxies, the total number of galaxies formally diverges.

Considering the component of hydrogen that correlates with optical
galaxy position, the average hydrogen mass per resolution element is
$M_{HI}=\Omega_{HI} ~\rho_{crit} V = 1.2\times 10^{11} M_{\odot}$.  We
are therefore reporting the detection of an aggregate $2\times 10^{14}
M_{\odot}$ of HI-emissive gas.  This is more than the sum of all
previously detected hydrogen in the universe. It is this large mass
that allows the detection of the 21-cm flux at a luminosity distance
an order of magnitude further than the furthest previous HI detection.
This illustrates the impact of Intensity Mapping technique.